\documentclass{elsart}
\journal{Physics Letters A}
\usepackage{graphics}
\usepackage{epsfig}

\begin{document}

\begin{frontmatter}
  \title{Characterisation of the dynamical quantum state of a zero
    temperature Bose-Einstein condensate\\} 
  \author[IC]{J. Rogel-Salazar\thanksref{cor}}
  \thanks[cor]{Corresponding author, e-mail: j.rogel@ic.ac.uk}
  \author[Rochester]{S. Choi}
  \author[IC]{G.H.C. New}
  \author[Oxford]{K. Burnett}
  
  \address[IC]{Quantum Optics \& Laser Science Group, Department of
    Physics, Imperial College, London SW7 2BW, U.K.}  
  \address[Rochester]{Department of Chemistry, University of Rochester, P.O. RC
    Box 270216, Rochester, New York 14627-0216} 
  \address[Oxford]{Clarendon Laboratory, Department of Physics,
    University of Oxford, Parks Road, Oxford OX1 3PU, U.K.}
  
  \begin{abstract}
    We describe the quantum state of a Bose-Einstein
    condensate at zero temperature.  By evaluating the $Q$-function we
    show that the ground state of Bose-Einstein condensate under the
    Hartree approximation is squeezed. We find that multimode
    Schr\"odinger cat states are generated as the condensate evolves in
    a ballistic expansion.
  \end{abstract}
  
\end{frontmatter}

\section{Introduction}
\label{sec:intro}

The realisation of Bose-Einstein condensation in dilute gases has made
it possible to study the dynamics of quantum fields directly
\cite{anderson,davies}.  In this paper, we address one of the most
intriguing issues surrounding a Bose-Einstein condensate (BEC), namely
the nature of its quantum state.  Number states might seem to be the
natural choice given that atoms cannot be created or destroyed at
these energies. However, it is well known that, in an open
environment, particles may be added or removed from the condensate,
and this implies that the condensate is not in a pure Fock state.  On
the other hand, it has been suggested that there should be a highly
entangled state of the condensate plus the environment \cite{barnett}.
The coherence properties of the condensate have been demonstrated
experimentally, indicating that a coherent state might be the most
robust representation \cite{andrews,anderson2,hall}.  This point turns
out to be very important, as it is well-known that a coherent state
propagating through an amplitude-dispersive medium evolves into a
superposition of two coherent states \cite{yurke}. The generation of
these Schr\"odinger cat-like states is considered to be a general
property of nonlinear systems where dissipation effects are not very
large.  Recently, some schemes to create Schr\"odinger cat states in
BECs have been proposed \cite{dunninham1} and the effects of loss have
also been considered.

It has also been pointed out that the state of the condensate might be
squeezed, given that binary collisions between atoms play a very
important role in the description of BECs \cite{keith1} and are known
to give rise to squeezing \cite{lewen,dunninham,bolda}. We note that
the question of describing the quantum state of a trapped BEC was
addressed in reference \cite{dunninham} where a single mode
approximation was considered, based on a symmetry-breaking picture.
The authors showed that for repulsive interactions the state is number
squeezed.  In this paper, we use a multimode description in the
Hartree approximation. In particular, we show how the state evolves
under ballistic expansion, which is important as this is the basis of
many experiments involving the condensates
\cite{andrews,anderson2,hall}.

To study the evolution of the quantum state of a BEC, we have used the
Heisenberg equation of motion obtained from the many-body Hamiltonian
describing the condensate, and used the Hartree approximation, under
which the $n$-atom wavefunction is written as a product of $n$ single
atom wavefunctions. This results in a nonlinear Schr\"odinger equation
for the condensate wavefunction, or the well-known Gross-Pitaevskii
equation (GPE). The GPE has proved to be a very useful tool for
studying the condensate dynamics, and its predictions are in very good
agreement with experimental results at low temperatures
\cite{jin1,mewes}. It is important to note that the GPE can be
understood in terms of either coherent states or number states. Here,
we make use of a number state description in order to obtain an
analogue of the scenario presented in \cite{lai} for the case of
light. We shall take ``snapshots'' of the wavefunction at various
times and analyse them using the $Q$-function, which is a
quantum-mechanical phase space function. We also want to mention that
we are treating the condensate in a zero temperature regime; this
means that we are basically referring to the predominant mechanisms
occurring in the ground state of the condensate. At finite
temperature, the mechanism is affected by the presence of excited
states and therefore the dynamics are changed \cite{rogel}.

The paper is organised as follows: in Section \ref{sec:qfunc}, we
present the Hamiltonian that describes the BEC and solve the equation
of motion in the Schr\"odinger picture using the Hartree
approximation. In Section \ref{sec:qstate}, we calculate the
$Q$-function for the condensate.  The evolution of the state of the
condensate in a ballistic expansion is then presented, and the
generation of Schr\"odinger cat-like states is observed. This is
relevant in studies of output couplers or atom lasers. Finally we
conclude in Section \ref{sec:conclusion}.

\section{Q-Function in the Hartree approximation}
\label{sec:qfunc}
The phase space representation is useful for visualising the evolution
of a quantum state, as the statistics of the state can easily be
described using the quasiprobability distribution functions
\cite{schleich}. In this case, we have made use of the $Q$-function
which has the nice property of being always positive.  This function
can be considered to describe probability densities, and has the
further advantage of being readily measurable by quantum tomographic
techniques \cite{bolda,mancini}. The first experimental measurement of
quantum states was reported by Smithey {\it et. al.} in 1993
\cite{smithey}.  Since then, several proposals to reconstruct the
quasiprobability distributions of a quantum system, such as the Wigner
function, have been made. For instance, Kurtsiefer {\it et. al.}
\cite{kurtsiefer} reported a technique to measure the Wigner function
of an ensemble of helium atoms in a double-slit experiment.  Moreover,
a proposal has been made to measure the quantum state of a BEC in an
atomic interferometer based on Raman transitions \cite{bolda}, where
the $Q$-function can be measured directly. The way of obtaining the
quasiprobability function is analogous to doing unbalanced homodyne
detection of an optical signal.

The $Q$-function of a pure quantum state $|\Psi\rangle$ is defined as
\begin{equation}
  \label{eq:qpure}
  Q(\alpha_r,\alpha_i)\equiv\frac{1}{\pi}|\langle\alpha|\Psi\rangle|^2,
\end{equation}
where the two real numbers $\alpha_r$ and $\alpha_i$ describe the
coherent state $|\alpha\rangle$. Therefore, the quasiprobability
distribution depends directly on these two parameters, which span the
phase space. An alternative form of equation (\ref{eq:qpure}), leads
the $Q$-function to be defined by the diagonal matrix elements of a
density operator $\hat\rho$ in a pure coherent state $|\alpha\rangle$:
\begin{equation}
  \label{eq:density}
  Q(\alpha_r,\alpha_i)\equiv\frac{1}{\pi}\langle\alpha|\hat\rho|\alpha\rangle.
\end{equation}

Clearly, the $Q$-function is characterised by being always positive
and normalised to unity.  Thus, it can be regarded as describing
probability densities, and the numbers $\alpha_r$ and $\alpha_i$ play
the role of conjugate variables, eg number and phase.

We shall start by describing the system in terms of the many-body
interacting Hamiltonian
\begin{eqnarray}
  \label{eq:m-bH}
  \hat H &=& \int {\rm d}^3{\bf r}\hat\phi^\dag({\bf r},\tau)\left[-\frac{\hbar^2}{2m}\nabla^2_{\bf r}+V_{trap}({\bf r})\right]\hat\phi({\bf r},\tau)\\
  & &+\frac{1}{2}\int {\rm d}^3{\bf r}{\rm d}^3{\bf r}'\hat\phi^\dag({\bf r},\tau)\hat\phi^\dag({\bf r}',\tau)\hat V({\bf r}-{\bf r}')\hat\phi({\bf r},\tau)\hat\phi({\bf r}',\tau),
\end{eqnarray}
where $\hat\phi^\dag({\bf r},\tau)$ and $\hat\phi({\bf r},\tau)$ are
the boson field creation and annihilation operators respectively,
and $\hat V({\bf r}-{\bf r}')$ is the two-body interatomic
potential.  For a contact interaction potential, the normalised
Heisenberg equation of motion for the operator $\hat\phi$ is given
by
\begin{equation}
  \label{eq:normGPE}
  {\rm i}\frac{\partial \hat\phi}{\partial t}=\left[-\nabla^2_{\bf r}+V_{trap}({\bf r})+\frac{nU_0}{\hbar\omega_{trap}}\hat\phi^\dag\hat\phi\right]\hat\phi.
\end{equation}

In the Schr\"odinger picture, equation (\ref{eq:normGPE}) takes the
form
${\rm{i}}\hbar\frac{{\rm{d}}}{{\rm{d}}t}|\Phi\rangle=\hat{H}|\Phi\rangle$.
The solution can then be found by expanding the state vector
$|\Phi\rangle$ in a Fock space
\begin{equation}
  \label{eq:fock}
  |\Phi\rangle=\sum_n A_n |n;t\rangle.
\end{equation}

The solutions of the equation are then number states $|n;t\rangle$. In
the Hartree approximation, it is possible to write the number states
as
\begin{equation}
  \label{eq:numstates}
  |n;t\rangle=\frac{1}{\sqrt{n!}}\left(\int {\rm d}x\Psi_n(x,t)\hat\phi^\dag(x)\right)^n|0\rangle,
\end{equation}
where $\Psi_n$ is a single particle wavefunction and $\hat\phi^{\dag}$
corresponds to the creation operator of the field.

This implies that each particle in the system experiences the same
potential. Therefore, the single-particle function $\Psi_n$ satisfies
the GPE with a scaled nonlinearity proportional to the number of
particles \cite{lai}.

If we superimpose these states with a Poissonian distribution, the
coefficient $A_n$ takes the form
\begin{equation}
  A_n=\frac{\alpha_0^n}{\sqrt{n!}}\exp{\frac{-|\alpha_0|^2}{2}}
\end{equation}
so that the quantum state in the Hartree approximation is expressed as
\begin{equation}
  \label{eq:qstate}
  |\Phi\rangle=\sum_n\frac{\alpha_0^n}{n!}\exp{\frac{-|\alpha_0|^2}{2}}\left(\int {\rm d}x\Psi_n(x,t)\hat\phi^\dag(x)\right)^n|0\rangle.
\end{equation}

The $Q$-function given by equation (\ref{eq:qpure}) can be written in
terms of a reference coherent state $|\alpha,\{\Psi(x,t)\}\rangle$ as
\begin{equation}
  \label{eq:Q}
  Q(\alpha_r,\alpha_i)=|\langle\alpha,\{\Psi(x,t)\}|\Phi\rangle|^2.
\end{equation}

We can construct the reference coherent state
$|\alpha,\{\Psi(x,t)\}\rangle$ using the $n$-particle eigenstate for
the field with envelope $\Psi_{\bar n}(x,t)$, where
$\bar{n}=|\alpha_0|^2$ has the meaning of the average particle number
\begin{equation} 
  \label{eq:refcoh}
  |n,\{\Psi_{\bar n}(x,t)\}\rangle=\frac{1}{\sqrt{n!}}\left(\int {\rm d}x\Psi_{\bar n}(x,t)\hat\phi^\dag(x)\right)^n|0\rangle.
\end{equation}
Assuming Poissonian statistics, we obtain the many-body coherent state
\begin{equation}
  \label{eq:manycoh}
  |\alpha,\{\Psi_{\bar n}(x,t)\}\rangle=\exp\left({\frac{-|\alpha|^2}{2}}\right)\exp{\left(\alpha\int {\rm d}x\Psi_{\bar n}(x,t)\hat\phi^\dag(x)\right)}|0\rangle.
\end{equation}

Using equations (\ref{eq:qstate}), (\ref{eq:Q}) and
(\ref{eq:manycoh}), the $Q$-function can be written
\begin{equation}
  \label{eq:Qdyn}
  Q(\alpha_r,\alpha_i,t)=\exp{\left(-|\alpha_0|^2-|\alpha|^2\right)}\times\left|\sum_{n=0}^\infty\frac{(\alpha^*\alpha_0)^n}{n!}\left(\int {\rm d}x\Psi_{\bar n}^*(x,t)\Psi_n(x,t)\right)^n\right|^2.
\end{equation}

\section{Quantum state of a Bose-Einstein Condensate at zero temperature}
\label{sec:qstate}
Let us consider Gaussian wavefunctions to describe the ground state of
the trapped condensate
\begin{equation}
  \label{eq:gaussian}
  \Psi_n=\exp\left[-\left(\frac{x}{n-1}\right)^2\right]
\end{equation}
and assume that $\bar{n}$ is large so that the difference between
$\bar {n}-1$ and $\bar {n}$ can be neglected. The $Q$-function for
such a case is
\begin{equation}
  \label{eq:Qgauss}
  Q_{G}(\alpha_r,\alpha_i)=\exp{\left(-|\alpha_0|^2-|\alpha|^2\right)}\times\left|\sum_{n=2}^\infty\frac{(\alpha^*\alpha_0)^n}{n!}\left[\sqrt{\pi}\left(\frac{\bar{n}^2n^2}{n^2+\bar{n}^2}\right)\right]^n\right|^2.
\end{equation}

\begin{figure}[htbp]
  \begin{center}
    \includegraphics[width=6cm]{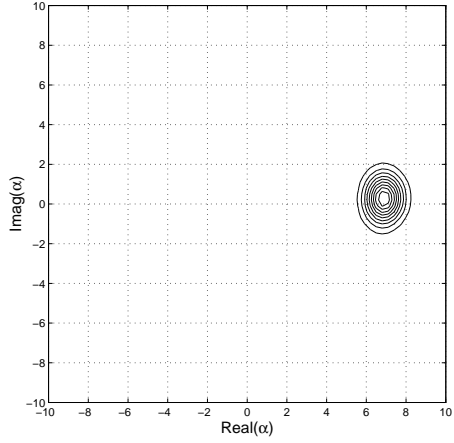}
    \caption{$Q$-function for a Bose condensed gas described by Gaussian functions.}
    \label{fig:Qgauss}
  \end{center}
\end{figure}

For an average number of particles $\bar{n}=50$ the $Q$-function
(\ref{eq:Qgauss}) is shown in Figure \ref{fig:Qgauss}.

\begin{figure}[htbp]
  \begin{center}
    \includegraphics[width=6cm]{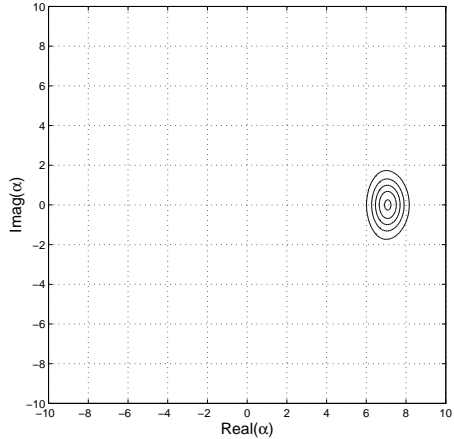}
    \caption{$Q$-function for a Bose condensed gas in the Thomas-Fermi approximation.}
    \label{fig:QTF}
  \end{center}
\end{figure}

A more realistic approximation to the ground state is given by the
Thomas-Fermi solution
\begin{equation}
  \label{eq:TF}
  \Psi_n=\frac{1}{\sqrt{n(n-1)U_0}}\left(\mu_{n}-V_{trap}\right)^{1/2},
\end{equation}
where the chemical potential $\mu_n$ is determined by the
normalisation of $\Psi_n$. The $Q$-function is then expressed as
\begin{eqnarray}
  Q_{TF}(\alpha_r,\alpha_i)&=&\exp{\left(-|\alpha_0|^2-|\alpha|^2\right)}\times\nonumber\\
  & & \left|\sum_{n=2}^\infty\frac{(\alpha^*\alpha_0)^n}{n!}\left\{\frac{4}{3}\mu_{n}\left[\left(\mu_{n}-\mu_{\bar{n}}\right){\rm F}\left(\sqrt{\frac{\mu_n}{\mu_{\bar{n}}}},\sqrt{\frac{\mu_{\bar{n}}}{\mu_n}}\right)\right.\right.\right.-\nonumber\\
  & & \left.\left.\left.\left(\mu_{n}+\mu_{\bar{n}}\right){\rm E}\left(\sqrt{\frac{\mu_n}{\mu_{\bar{n}}}},\sqrt{\frac{\mu_{\bar{n}}}{\mu_n}}\right)\right]\right\}^n\right|^2,\label{eq:QTF}
\end{eqnarray}
where ${\rm F}(\varphi,k)$ and ${\rm E}(\varphi,k)$ are incomplete
elliptic integrals of the first and second kind respectively
\cite{byrd}.  For a harmonic trap potential
$V_1=\frac{1}{2}m\omega_{trap}^2r^2$, the chemical potential is
given by $\mu_{n}=\left(\frac{3n(n-1)U_0}{8}\right)^{2/3}$. The
$Q$-function for the same average number of particles as in the
previous case is plotted in Figure \ref{fig:QTF}.

From Figures \ref{fig:Qgauss} and \ref{fig:QTF}, it is clear that
the ground state of the condensate is very close to a squeezed
state, since the $Q$-function has different widths in the $\alpha_r$
and $\alpha_i$-directions. On the one hand, the squeezing is
generated by the nonlinearity inherent to the system. This can be
explained in terms of a superposition of different numbers of
particles with different phases, creating a deviation from the
classical phase. On the other hand, we know that the coherent state
interpretation arises from the robustness of these states in the
presence of interactions between the condensate and its environment
\cite{barnett}. It is well known that coherent states are a
particular case of a more general class of minimum-uncertainty
states, namely the squeezed states. It has been shown that the
propagation of a coherent state in a nonlinear medium evolves into a
superposition of two coherent states \cite{yurke}, while the
quantum-mechanical superposition of two coherent states has been
shown to lead to a transition from a Poisson distribution to a
sub-Poissonian one and squeezing \cite{schleich1}.  In this sense,
the robust description of the quantum state still applies, bearing
in mind that the existence of the interactions between atoms yields
squeezing of the quantum field and, as a result, entanglement
between the atoms is also produced \cite{keith1}. We note that while
a single coherent state does not evidence squeezing, a considerable
amount is exhibited in a superposition of even just two coherent
states.

This mechanism of achieving squeezing is the predominant one in the
case of the ground state, without the presence of highly populated
excited states. When we consider the presence of a thermal cloud, we
can treat the system using the Bogoliubov procedure and analyse the
behaviour of the collective excitations. It has been shown that
squeezed quasiparticle excitations can be produced when considering
the generation of correlated pairs by the interaction of an excited
state with the ground state \cite{rogel}. Thus, we can talk about
two different aspects of nonclassicality in the system, the
predominant one for the ground state at zero temperature, and the
prevalent one in the presence of excited states.

\begin{figure}[htbp]
  \begin{center}
    \includegraphics[width=6cm]{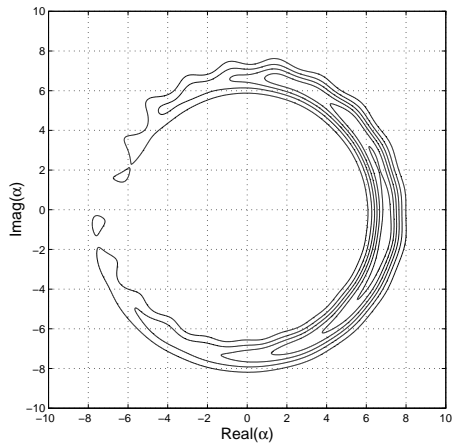}
    \caption{$Q$-function for the dynamical evolution of the system at $2\times 10^{-3}$ s.}
    \label{fig:free2}
  \end{center}
\end{figure}

\begin{figure}[htbp]
  \begin{center}
    \includegraphics[width=6cm]{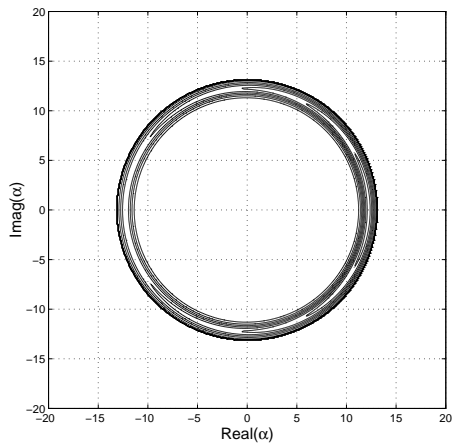}
    \caption{$Q$-function for a typical thermal state}
    \label{fig:thermal}
  \end{center}
\end{figure}

It is of particular interest to see how the $Q$-function behaves
once the condensate is released from the trap as when the atomic
cloud is allowed to expand ballistically \cite{anderson}, or in some
output coupling schemes for atom lasers \cite{helmerson}. The
Thomas-Fermi solution was used as an initial condition to solve the
GPE after the trapping potential is switched off and the
$Q$-function (\ref{eq:QTF}) was then calculated from the dynamical
evolution of the wavefunctions.  We expect that once the trapping
potential is switched off, the condensate will rethermalise.
Comparing the quantum state of the condensate, for a trap frequency
$w_{trap}=2\pi\times 100$ Hz, at $2\times 10^{-3}$ s (shown in
Figure \ref{fig:free2}) to the $Q$-function for a thermal state
(Figure \ref{fig:thermal}) makes it clear that rethermalisation is
indeed taking place.

\begin{figure}[htbp]
  \begin{center}
    \includegraphics[width=6cm]{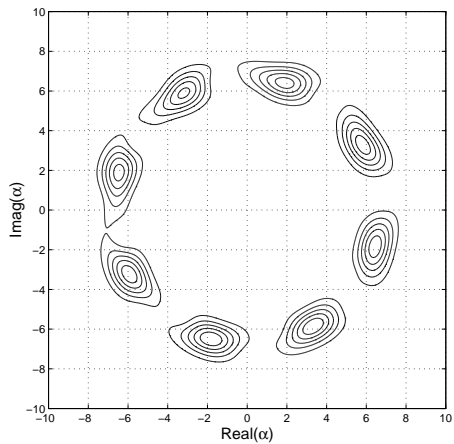}
    \caption{$Q$-function for the dynamical evolution of the system at $5\times 10^{-3}$ s.}
    \label{fig:free5}
  \end{center}
\end{figure}

\begin{figure}[htbp]
  \begin{center}
    \includegraphics[width=6cm]{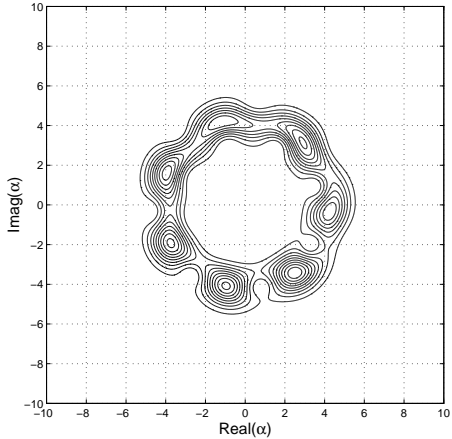}
    \caption{$Q$-function for the dynamical evolution of the system at $1\times 10^{-2}$ s.}
    \label{fig:free10}
  \end{center}
\end{figure}

A very important aspect that has to be considered is the finite
lifetime of the condensate, which is at most 10-20 s.  The reason
behind this, is that the thermal cloud is not the only source of
decoherence. The condensate loses atoms because of Rayleigh
scattering, external heating, and three-body decay. Information
about the quantum state of the condensate is carried by the atoms
that escape from it so that, due to interactions of the system with
the environment, a quantum superposition is turned into a
statistical mixture. From the processes mentioned above, three-body
recombination is the most important one \cite{burt,fedichev}. In an
experiment carried out by Stamper-Kurn {\it et. al.} \cite{stamper},
the measured loss rate per atom turned out to be $4/s$ for $N=10^7$
atoms. This rate scales as $N^3$, where $N$ is the number of atoms
in the condensate.  Considering a condensate with $N=10^4$, just one
atom is lost per second; the decoherence time is 1 s. Bearing this
fact in mind, it is perfectly sensible to ask whether, after opening
the trap, it is still possible to see a Schr\"odinger cat-like
state.  When the states evolve further, a multi-component structure
develops as shown in Figures \ref{fig:free5} and \ref{fig:free10},
at $5\times 10^{-3}$ s and $1\times 10^{-2}$ s, respectively. In
both figures, the quasiprobability function demonstrates that the
quantum states obtained correspond to Schr\"odinger cat states like
the ones generated in Kerr media \cite{fu,miranowicz}.  Moreover, we
also see that the cat states have a different number of components
at different times in the evolution, in a close analogy to the
mechanism discussed by Yao for the case of a radiation field
propagating in a nonlinear medium \cite{yao}. It has been shown that
the quantum-mechanical superposition of states can yield a state
that exhibits squeezed fluctuations \cite{knight,schleich1}. This
suggests that, even in the presence of losses due to three-body
recombination, it would still be possible to observe Schr\"odinger
cat-like states.
 
\section{Discussion}
\label{sec:conclusion}
In summary, we have described the ground state of a Bose condensed gas
in the Hartree approximation in a phase-space representation using the
$Q$-function. We have confirmed that the ground state of the
condensate is in a squeezed state due to the presence of interactions
between the atoms. We have also studied the dynamical evolution of the
condensate in a ballistic expansion.  The simulations show that
immediately after the condensate is released from the trap the system
rethermalises, with the quantum state being still close to a squeezed
state.  Continuing the evolution without the trapping potential, we
obtain a multimode Schr\"odinger cat-like structure. In general, the
``longevity'' of these states is very small due to decoherence.
However, the time scales of the calculations suggest that it might
still be possible to observe them.  The generation of nonclassical
states by the time evolution of an initial coherent state under the
influence of some nonlinear dynamics has been studied in the past for
the case of radiation fields \cite{yurke,fu,miranowicz,yao}.  In the
case of a BEC the nonlinearity arises from binary correlations between
the atoms, producing the effective interactions between them. We point
out that this process basically refers to the ground state, as the
presence of excited states changes the predominant mechanisms of
achieving squeezing.

An issue that remains to be addressed is the one related to the
experimental reconstruction of the quantum state of the condensate. We
have mentioned that it is possible to use quantum tomographic
techniques to measure the $Q$-function \cite{bolda,mancini}. As an
alternative to these techniques, a scheme using projection synthesis
was proposed by Baseia and co-workers \cite{baseia}. The scheme
follows the idea of measuring the probability distribution of a state
$|f\rangle$ to have an observable $\theta$
\begin{equation}
  \label{eq:probability}
  \langle f|\theta\rangle\langle\theta|f\rangle=P_N(\theta)=\frac{1}{2\pi}\left|\sum_{n=0}^Nc_n \exp(-{\rm i}n\theta)\right|^2,
\end{equation}
where the expectation value is replaced with the probability density
by setting $c_n$ to zero for $n>N$, with large $N$. The probability
density is proportional to the expectation value of the projector
$\hat \pi=\kappa |\theta\rangle\langle \theta|$ where $\kappa$ is a
positive constant. If the field is in a mixed state $\hat \rho$,
equation (\ref{eq:probability}) can be written as
\begin{equation}
  \label{eq:rho}
  P_{\rho}(\theta)={\rm Tr}(\hat \rho \hat \pi)=\langle\theta|\hat \rho|\theta\rangle.
\end{equation}

It has been shown \cite{barnett1} that by specifying a suitable
reference state $|B\rangle_b$ in an experimental setup where a state
$|\psi\rangle_a$ is coherently mixed in a beam splitter with the
reference state $|B\rangle_b$, we can obtain the required probability
(\ref{eq:rho}).

In the case of the quasiprobability function $Q$, we would require a
projector $\hat \pi=\kappa |\alpha\rangle\langle\alpha|$ so that its
substitution in equation (\ref{eq:rho}) yields
\begin{equation}
  \label{eq:projQ}
  P_{\rho}(\alpha)={\rm Tr}(\hat \rho\hat\pi)=\langle\alpha|\hat\rho|\alpha\rangle=Q(\alpha_r,\alpha_i)
\end{equation}

The form of the reference state that leads to the required projector
is given by \cite{baseia}
\begin{equation}
  \label{eq:reference}
  |B\rangle_b=C\sum_{k=0}^Nb_k|k\rangle_b,
\end{equation}
where $C$ is a normalisation constant and
\begin{equation}
  b_k=\frac{2^{-N}}{C}\left(\begin{array}{c}
      N\\k\end{array}\right)^{-1/2}\exp\left(\frac{{\rm i}k\pi}{2}\right)C_{N-k}^*.
\end{equation}

It is important to mention that this scheme involves projections onto
truncated coherent states, which is a good approximation for large
$N$. In other words, it works well for the case $N\gg\bar{n}$, where
$\bar n$ is the average particle number of the synthesised coherent
state. The construction of the reference state (\ref{eq:reference})
would be the main experimental challenge in this scheme. However, one
might able to generate such a state using the scheme proposed
in \cite{dakna} where the authors show that it is possible to prepare
an arbitrary (finite) superposition of Fock states by applying a
well-defined succession of the displacement and creation operators.
Experimentally a sequence of beam splitters at which conditional
measurements are performed is required. In the case of BEC, this may be
achieved by using repeated interference of condensates with different
hyperfine states, similar to the Gedanken experiment proposed by
Castin and Dalibard in reference \cite{castin} where atoms leak from
two trapped condensates and are detected in the output channels of a
50-50 beam splitter. Exact procedure for such state preparation is the
subject of our future work

This work has been supported by CONACyT, EPSRC, the Royal Commission
for the Exhibition of 1851 and the EU.

\end{document}